# Some General Remarks on Private Property

Adnan N. Alabbar[1] and Walter E. Block[2].

[1] Texas A&M University, Department of Physics and Astronomy, College Station, TX 7843-4242 USA, and [2] Loyola University, 6363 St. Charles Avenue, Box 15, Miller Hall 318, New Orleans, LA 70118, USA: Corresponding Author: Adnan.alabbar94@tamu.edu.

## Abstract.

Private Property is one of the central institutions of civilized society. We first consider its social, legal, and economic aspects. We then follow the Lockean tradition by focusing on a specific procedural definition: Homesteading is the acquisitive act of first using an object that is initially unowned. The ontology of concrete objects and the nature of their uses determine how objects may be acquired. In this article, we address the open-texture problem in the definition of property, then provide the necessary conditions for an object to be property in the Lockean Scheme.

> *"We shall never save civilisation as long as civilisation is our main object. We must learn to want something else even more."*
>
> — C. S. Lewis (Mere Christianity)

# 1. Introduction.

The institution of property has existed as far back as recorded history allows. Upon examining the meanings of the term throughout the past, one finds that it has always been ill-defined: For what might *it* be such that it encompasses human bodies, animals, plants, land, buildings, machines, ideas, techniques, names, reputations, consumables, processes, relations, promises, and even radio frequencies? It seems to be an overly capacious category for such a heterogeneous collection of objects, and a notion this amorphous and broad must be clarified before it can be useful.

Furthermore, throughout history, property has been said to be owned by individuals, families, groups, communities, corporations, governments, and sometimes even by all living human beings. What does it mean to say that some particular thing is mine, or thine, or ours, or theirs, or everyone's, and what rights and duties does such ownership entail? Moreover, can some things be owned partially and others entirely, and what limits may be placed on ownership, and on what grounds?

Additionally, by what mechanisms can modern man still acquire property, and how does one lose property? Are there any clear-cut standards that can be used to fairly attribute (or distribute) objects to individuals or institutions? To give a relevant example on the abandonment of the Lockean homesteading paradigm of original acquisition: Two centuries ago, one could own any patch of previously unowned land by building a house on it, or a field by fencing it for his cattle to graze, or extracting oil from it. This state-of-affairs is unfathomable today, as almost every state would not allow it (Alabbar, 2023), and even the citizens themselves might find that objectionable. Across the world, property questions lie at the heart of many of our most heated ethical and legal disputes: the ownership of handguns is contested between the right to personal safety and the right to self-defense; arguments over abortion now turn on a woman's ownership of her own body; and conflicts over the collection and sale of personal data hinge on who owns our digital traces; and so on.

These questions have been relevant to scholars for time immemorial. They are still hotly debated by philosophers, economists, legal theorists, and ethicists because, in each field, property is considered from different angles, and every scholar has their own dispositions.

From these diverse contributions from disparate times, cultures, and climates of thought, there still is no objective account of property, and there very well might never be one that most intellectuals can agree on, seeing that each theory leads to entirely different expectations, normative realities, and systems of power.

For example, a central fault line between classical liberals and communists is their stance on private property rights. Both might accept a thin characterization of "property rights" as the normatively justified locus of control over resources: i.e., who may use them, exclude others, and transfer them. But that apparent agreement immediately reopens the decisive question: what counts as "justified," and by which ethical standard? Moreover, jurists typically work with a broader, historically layered conception of property (not merely control over objects but a complex bundle of claims, powers, and duties), and they are often reluctant to abandon that inherited framework.

However, our language has been refined over the ages to allow us to probe the concept with ever more precision and alacrity, and this progress will help us understand what property is, and what it cannot be. There is a hope that, for better or for worse, there can be a definition that is objectively more useful, and therefore, will be more readily accepted by most parties: It is this definition that we seek to find, if it can be formulated.

Nevertheless, it should be borne in mind that there can never be a definition that will be accepted by all. Three issues become immediately salient: (1) The definition of property (i.e., what is legitimate property in the eyes of the law, or what may be physically possessed and appropriated without committing any form of aggression) is drawn from an ethical theory, and people differ in their ethical beliefs; (2) State officials—bureaucratic, legislative, executive, and judicial—as well as other political elites, have an incentive to redefine property in manners that are lucrative to them, and conducive to their particular goals; (3) Due to a history that extends to even before our earliest written memories, many things were called property so as to extend the ontological concept itself, and contingencies unimaginable to people before the 20th century are now said to be appropriable (e.g. Internet domains and cryptocurrency).

Three generally pertinent methods of defining property seek to affect different outcomes. One method is concerned with *efficiency*: It asks the question, what should be included in the set of appropriable objects, and how should property rights be defined, and (in cases of disputes) reallocated, such that we can achieve greater economic prosperity and enhance the structure of production.[1] Patents and intellectual property rights are often included under this rubric (Plant, 1934A; 1934B; Landes and Posner, 2003; Palmer, 1990; Spooner, 1884).

[1] Harold Demsetz and Ronald Coase are important champions of this line of thought. Consider, for example, Demsetz (1979): "Efficiency calls for a high degree of stability in property rights definitions, but it does not necessarily forbid all involuntary reassignment, especially when high exchange cost or free-rider type problems reduce the efficacy of allocations through the market." For a discussion and a challenge to their paradigm, see (Block, 1977; 1995; 2000; 2010) and for Demsetz's and Brooks' responses, see (Demsetz, 1979; 1997; Brooks, 2007). Thomas Sowell (1996) gives another interesting argument from efficiency, which we will not discuss here.

Another, with *justice*: Some economists or legal theorists dating back from Aristotle and the Stoics, extending to St. Aquinas and the scholastics, and till the present day, consider the definition of property to fall under natural rights theory (Rothbard, 1982; 2002; Buckle, 1993); others may want to redefine property in ways to limit citizens' political control, reduce inequality, or relieve poverty, famine, and suffering by, for example, asserting that all people may be entitled to subsistence.[2] These arguments are ethical in nature.

A third method is ontological. It asks what sorts of things can, in principle, be owned at all, and what kind of relation must obtain between a person and an object for ownership to arise. Locke (2014) was perhaps the first to articulate what is now called the 'homesteading' theory of ownership: whoever first "mixes" his labor with a previously unowned resource thereby acquires a property right in it. On this view, appropriable objects are limited to those that can be homesteaded in this way. It is still disputed, however, whether the "use principle"—that is, which uses

[2] Thus, Sen (1988):
"If a set of property rights leads, say, to starvation, as it well might, then the moral approval of these rights would certainly be compromised severely," or quoted therein, the Indian constitution grants "the right to an adequate means of livelihood."
As for Rawls (2001, p. 114), he differentiates between higher-order goods (means of production) and lower order goods (consumption goods) in his analysis:
"Among the basic rights is the right to hold and to have the exclusive use of personal property. One ground of this right is to allow a sufficient material basis for personal independence and a sense of self-respect, both of which are essential for the adequate development and exercise of the moral powers. Having this right and being able effectively to exercise it is one of the social bases of self-respect. Thus this right is a general right: a right all citizens have in virtue of their fundamental interests. Two wider conceptions of the right to property are not taken as basic: namely,

(i) the right to private property in natural resources and means of production generally, including rights of acquisition and bequest;

(ii) the right to property as including the equal right to participate in the control of the means of production and of natural resources, both of which are to be socially, not privately, owned."
For a response to Sen and Rawls (1971), see (Nozick, 2013, Ch. 7).

suffice to homestead an object—extends to nonphysical uses, such as the mental conception of an idea, or only to physical uses, such as employing the object in the process of production. However, we note that it becomes an empirical question to determine whether something is property.

An opposing ontological account is legal positivist: On this view, property is not a natural relation that develops from the interaction of the people, but a legal abstraction set by legislators and jurists. Bentham (1938, pp. 308–309) starkly makes the point: "property is entirely the creature of law." In consequence, there is no physical "trace" in an object that could reveal ownership as a matter of fact; ownership is an intelligible relation only within a framework of rules that defines and secures claims of use, exclusion, and transfer. Hence Bentham concludes that property and law stand or fall together, so that "before the laws, there was no property," and without law, property "ceases."

These three ways of defining property—efficient, ethical, and ontological—differ in what they aim to justify, but they all presuppose that private property is a real and legitimate social institution with determinate effects that need to be studied and engineered.

The ontological aspect (i.e., the nature of private property, per se) will be our focus in this article. In what follows, we turn from these justificatory and conceptual questions to (§2) the social, legal, and economic roles that private property in fact plays, beginning with its central functions in coordinating human action. In (§3), we discuss the ontological qualities of private property as a prelude to a definition. In (§4), we argue that the openness of the concept of private property is limited by considering necessary and sufficient conditions, and summarize and then conclude in (§5).

# 2. Social, Legal, and Economic Aspects of Property.

## (2.1) The Six Social Functions of Private Property.

Ownership has always been an *active* enterprise. Private property serves several social functions that facilitate human actions and are unlikely ever to become obsolete. It takes on an almost magical dimension in our social lives: Each of us carries an image of how the world ought to be, yet our attempts to realize it are immediately checked by our ignorance of how others will respond or what the circumstances will be. As Hayek (1988) put it, "the curious task of economics is to demonstrate to men how little they really know about what they imagine they can design." Many social thinkers treat economics as if it were the analysis of a linear system, in which their task is to compute the system's response to our stimuli (Sowell, 1996). Our attempts to affect any change in the world, our attempts to organize it, will necessarily produce *some* results that cannot be predicted beforehand. This is not because there are no social 'laws,' the very issue disputed in the Menger–Schmoller debate (Louzek, 2011), but because such laws can at best yield a general, never a particular, picture of the system's response. A general system of private property offers the highest practical protection against aggressive interference, and thus a modest but indispensable degree of predictability and continuity for our efforts to bring about these changes. But since there are no *constants* in the domain of the social sciences allowing us to deal with the world ideally, such a system can only take the form of a legal framework that mitigates, rather than eliminates, the non-ideal character of the world.

A second frustration concerns what Nozick (2013, pp. 57–58) calls the boundary-crossing of others' rights: to satisfy our own aims, we require freedom from interference, and, conversely, we must refrain from interfering with others' legitimate efforts to meet their own aims. Property rights, together with the Nonaggression Principle (Rothbard, 2002), circumscribe these boundaries.

We may separate the functions of private property and their associated rights into six central ones. This first function is the *allocation of control* (Block, 2019; Rothbard, 2009B, pp. 91 - 93; Merrill, 1998). Since production precedes consumption, and given that humans will always have ends to satisfy which require means that necessarily deploy scarce resources, the owner of a given resource determines which ends those scarce means (with alternative uses) will serve, and therefore what is to be produced, and in which manner, in order to be eventually consumed. (This is especially important for rivalrous goods.) The owner can then use their property to choose which goods and services to be produced, at what levels of sustainability and efficiency, engage in trade, decide whom to trade with and on what terms, and enjoy the resulting fruits of their production. This function gives the owner a limited sphere of control, excluding all others, for the satisfaction of his ends.

The second function is that of *incentive* (Demsetz, 1967; Alchian and Demsetz, 1973; Sowell, 1996). Private property concentrates both the benefits and the burdens of decision-making in identifiable owners. If property suddenly became an illegitimate institution, or if a person's assets were confiscated, declared 'public' or communal, or forbidden from ownership by anyone, the profits from their productive use would either go unharvested or be diluted across very large groups, so that no individual would have a strong reason to bear the costs of maintaining, repairing, or replenishing the underlying means of production, and many desires go unsatisfied. In the case of central control, responsibility for upkeep would be diffused to distant agencies that are less interested in careful stewardship, or to no one in particular, and no single decision-maker would have both the motive and the authority to choose sustainable and efficient methods of production (Foss et. al., 2020). In such a regime, the incentive to protect shared resources, including the environment, is weakened.

Conversely, the third function concerns *positive and negative externalities and conflict resolution* (Mises, 1998, pp. 650 - 652; Rothbard, 1982; Alchian, 1965). Private property makes it possible to assign responsibility for negative externalities to a determinate owner, whose use of an asset is constrained by the prohibition on trespassing others' boundaries; that owner can then be held liable for damages and required to indemnify those harmed if physical injuries or damages occur. Losses, however, are tricky.[3] At the same time, the same property structure allows owners to devise ways of earning revenue from the positive externalities generated by their assets. Disputes over incompatible uses of the same resource can then be resolved by appealing to the underlying property titles—who owns what, and what contractual rights or easements exist—rather than by resorting to ongoing threats, bargaining under duress, or open violence. In this way, property rights help to establish clear boundaries between individuals and clarify when defensive force or legal action is permissible.

[3] Thus, Block (2019):
"Confusion abounds on the issue of whether property rights concern the value of physical things or, instead, it is the physical things themselves which are of value. It is thus necessary to clarify why the common notion of property rights as extending exclusively to physical things is indeed correct, and why the notion of property rights in values is flawed. [...] It is easily recognized that every action of a person may alter the value (or price) of another person's property. If A enters the labor or the marriage market, this may impair B's value in these markets. And if A changes his relative evaluation of beer and bread, or if A decides to become a brewer or a baker himself, this may change the property values of the—other—brewers and bakers. According to the view that value- impairments constitute rights violations, it follows that A's actions may represent punishable offenses. Yet if A is guilty, then B and the brewers or bakers in turn must be entitled to defend themselves against A's actions. Their right to defend themselves can only consist in their (or their agent) being permitted to physically attack or restrict A and his property: B must be entitled to physically bar A from entering the labor or marriage market; and the brewers or bakers must be allowed to physically hinder A from spending his own money as he pleases, for example, from using his own possessions for the operation of a brewery or bakery. Based on this theory, the physical damaging or restricting of another person's property use obviously cannot be said to constitute a rights violation. Rather, physical attacks and physical restrictions on the use of private property then have to be classified as lawful defenses. On the other hand, suppose that physical attacks and physical property restrictions constitute rights violations. Then B and brewers or bakers are not allowed to defend themselves against A's actions. For A's actions—his entering the labor or marriage market, his changed evaluation of beer and bread, and his opening of a brewery or bakery— affects neither B's bodily integrity nor the physical integrity of other brewers' or bakers' property. If they engage in physical resistance against A's actions nonetheless, then the right to defense rests with A. In this case, however, it cannot be considered a rights violation that a person's actions impair the value of another person's property. No other, third alternative exists."

The fourth function is the facilitation of economic calculation (Mises, 2012; Rothbard, 1991). Only on the basis of private property can we have economic calculation and market-based price formation, because only where ownership claims are clearly defined and transferable can markets generate the relative prices that guide rational economic calculation. In trade, sacrificing one's property, say money, in exchange for another's products, determines the demonstrated preferences of both, shows where each item was in each person's value scale, and ultimately determines the market prices of each (§2.4).

The fifth function consists of *local governance and rule-setting* (Ostrom, 2009). Private property allows (partial) owners to set rules regarding how the owners, and others, may act on a given object, or within a given area or system, so that each owner can establish a local order—standards of conduct, access, and use—within the broader framework of general rights. For example, different households tolerate different levels of cleanliness or untidiness, but each room can differ in these levels within a reasonable limit agreed by the group; medical clinics may specialize in particular conditions and adopt admission criteria based on expertise, capacity, or contractual arrangements; and schools impose their own rules of conduct, uniforms, and grading policies. It is often less efficient to centralize all such rule-setting in a single authority, especially when an object is owned by a group. Decentralized ownership allows different institutions to apply different rules, to learn from one another's successes and failures, and to better match local conditions and preferences. This function pertains especially to jointly held property and thus differs from the first function, in which control cannot be exercised on equal terms by all participants; as a result, co-owners are subject to specific rules. Some authors associate this governance role with the institutional core of a minarchist society. Still, we emphasize that such arrangements are legitimate only insofar as they rest on unanimity among those subject to the rules, since citizens do not own aliquot shares in their governments.

Finally, the sixth function of private property is that *it grants owners the freedom of association* (Mises, 2009). Private property underwrites a significant degree of personal independence by allowing individuals to associate with whomever they wish, thereby making it possible to specialize, to form and dissolve cooperative arrangements, and to decide for themselves on whom they will depend in production and exchange, rather than having these relationships imposed by collective decision.

Together, these inter-related functions make private property a permanently active institution: it allocates control over scarce means, properly structures incentives, sustains 'long-run' production, assigns responsibility for harms and benefits, allows people to choose whom to associate with and what rules to set locally, and furnishes the informational signals (market prices) required to coordinate plans in a world of genuine uncertainty about future preferences and outcomes (Mises, 2008) and irreducibly dispersed, (and perhaps tacit), knowledge (Hayek, 1945).[4]

Some scholars deploy property as a tool in the service of further goals—such as promoting social equality or enforcing particular patterns of distributive justice (Rawls, 1971; 2001), articulating a "social-obligation" view that directs owners to use their holdings in ways approved by political or social authorities (Alexander, 2009), treating racial identity itself as a form of property (Harris, 1993), or construing some forms of ownership as constitutive of personhood and identity (Radin, 1982). From our perspective, however, these are not additional functions of private property at all, but either normative glosses on the six functions just outlined or descriptions of what happens when property rights are overridden or redefined; in any case, they are contrived and can be ignored. Once property is understood as a system of exclusive control over scarce physical resources, acquired by homesteading, contract, and restitution (Block, 2019, Part III), the principal social roles it plays are exhausted by the six functions above.

[4] We have not included the security or pleasure that ownership might grant because these two are subjective estimations of the owner, and we have focused on the objective social functions.

### (2.2) Some Social and Historical Aspects of Private Property.

Many historians believe that collective property (i.e., objects owned by a group, such as a tribe or family) preceded private property (i.e., objects owned by individuals). Features of shared or common property are thought to be the oldest in the human race. These evolved into a personal property system, perhaps with the condensation from the large tribe smaller family units (Hoppe, 2015, Ch. 1). Aspects of property can even be witnessed in the animal kingdom, suggesting that it might predate humanity.[5] Though its nature may differ across species, property is an essential social institution in all species that exhibit it.

---

[5] Jeffrey Stake (2004) provides evidence that many animals and birds show, to a simpler degree, signs of honoring the private property of their conspecifics, as well as recognizing their own through defending it. This not only suggests that private property is not only a human institution but that it might be a natural mechanism that precedes homo sapiens as a species.
Property can be seen in primates particularly clearly. To quote Currier (2017):

"Primates also distinguish between two fundamentally different types of ownership: communal property and personal property. The territory and natural resources of a group's homeland—

Suppose that an early human caught a fish or found a berry shrub. In that case, we imagine that he had to share it with other tribe members to be protected, or if, through repeated exclusion, he barred them from accessing what they thought to be the fair property of the group, he may well have been exiled from it, which would have meant certain doom for the tribesman. For a modern human, the concept of property is no less social: For how can one object be attributed to one person, except vis-à-vis others?

Ownership is not a meaningful action for a Robinson Crusoe stranded alone on an island. In owning something, there aren't any other people for rules to be set around that object. Furthermore, the ways we use property are socially interactive: A Robinson Crusoe might collect wood logs and twigs for immediate use, or to save them for future use, but all the other uses of wood that exhibit the exploitation of the law of comparative advantage would be lacking for him. Ownership is necessary for interpersonal, not autistic, exchange.

Private property introduces into social life an ethic of mutual non-interference. Each person may choose his own way of life and pursue his goals by means of his property; each household can adopt rules that differ

---

including sleeping trees, fruit trees, honeycombs, birds' nests, drinking places, and so on—are generally considered to be the communal property of the group as a whole, and any member of the group has a right to use them. But when a particular piece of fruit is picked, a tasty insect is captured, or a nest of branches is constructed in a sleeping tree, that nest or morsel of food becomes the property of the individual who gathered or built it—and it is rarely shared with others.

"Human societies all recognize these two types of property in the personal possessions that belong only to individuals versus the public territory shared by all members of the community (which in our society includes streets, roads, parks, and other public spaces). To these, humans have added a third type of property, the family possessions (especially food and dwellings) that are shared by the members of the family but not by the society as a whole."

For more on family possessions, see also (Hoppe, 2015).

from those of its neighbors, and so on. The private character of property allows individuals to specialize, cooperate on their own terms, and set their own rules, rather than live at the mercy of others' whims. Interferences such as theft, fraud, and murder can be challenged by individuals and by private firms specializing in adjudication, contract enforcement, insurance, protection, and related functions (Rothbard, 2009B, Book 2; Hoppe, 2009). A communist objection might be that such interferences would not arise without private property existing in the first place; our response is that the six functions of private property enumerated in (§2.1), which evolved spontaneously over the centuries, were necessary to solve problems of production and welfare, and that any realistic institutional order must confront the trade-offs they involve.

### (2.3) The Legal Functions of Private Property.

As is the case for most social institutions, all modern human societies have at least some conception of 'private' property that is deeply respected, and many laws are designed to protect it *from others*. And again, as with most social institutions, each society differs slightly in what it includes as property and, therefore, in what local laws protect and what they exclude or overlook, due to prevailing social practices.

Anthony M. Honoré points out, however, that certain rights persist in most conceptions of ownership. Such rights include immunity from arbitrary expropriation, the power to alienate the property or transfer it, and the liberty to consume or destroy it, choosing who to manage and use it, among other things (Honoré, 1961). These rights are shared and inscribed in the legal codes of most societies today.

This conception, which follows Hohfeld's (1913) analysis of the rights of "the owner of ordinary personal property 'in a tangible object',"

is called the Bundle of Rights theory.[6] Here, property *rights* are a set of privileges, duties, powers, disabilities, immunities, and liabilities toward individuals, some of which may be more important than others, and some can be replaced or even removed. Societies differ in what 'sticks' belong to such a bundle. But then, following the *Bundle of Rights* theory of property, legal property, and property rights, thus, are subjective: as some of the particular rights differ from one society to another, even if there exist some general and shared rights that do not.[7]

This might be favorable to some: on the one hand, there is greater freedom for legal systems to differ, and comparisons of their consequences can be organized. Commentators may examine the successes and failures of others through the ethical values they adopt and act accordingly. This applies to individuals as well as to societies (collections of individuals sharing traditions, cultures, languages, and geography).

Likewise, if society A adds stick X or removes stick Y from the bundle, and society B does not, we can study both legal systems and their consequences. This goes to show that the existence of a bundle, per se, is not only necessary, but utilitarian, and instructive for development and growth.[8] This helps societies learn from each other and evolve; this is opposed to a unified system where errors are often harder to detect because nothing radically different has ever been tried.[9]

---

[6] Some modern legal 'properties,' such as intellectual property or trademarks, are not tangible.

[7] In philosophical terms, the general quiddity (essence) is shared but the particular haecceity (thisness) isn't. The essence of the bundle may be kept intact, but if some sticks differ from one society to another, then it becomes subjective. This is not to deny that some versions thereof may be more *just* than others.

[8] This insight was perhaps first discovered by the Presocratic philosopher Democritus (Rothbard, 2006, p. 10): "Democritus's other major contribution to economics was his pioneering defence of a system of private property. In contrast to Oriental despotisms, in which all property was owned or controlled by the emperor and his subordinate bureaucracy, Greece rested on a society and economy of private property. Democritus, having seen the contrast between the private property economy of Athens and the oligarchic collectivism of Sparta, concluded that private property is a superior form of economic organization. In contrast to communally owned property, private property provides an incentive for toil and diligence, since 'income from communally held property

On the other hand, this poses a problem for anyone who believes that ethics and legal theory are *objective* sciences in that they attempt to discover laws that govern our social reality justly or optimally,[10] and that veering away from the proper bundle of sticks will unethically experiment on individuals—especially since these rights are often imposed on them.[11] This reminds us how important it is to try to find what the correct bundle is, if, indeed, there is one.[12]

Suppose that stick X allows individuals to encroach on some of the liberties of others. Will we let it for the sake of experimentation, knowing that some people's rights will be abridged? One famous example would be intellectual property rights. Kinsella (2001) considers this problem in greater detail: If I claim intellectual property rights to a patent for an invention, I can deny another individual the freedom to use the same blueprints and *his own property* to make the same device to which I hold a patent. This imposes a limit on his ownership that many economists and

---

gives less pleasure, and the expenditure less pain.' 'Toil,' the philosopher concluded, 'is sweeter than idleness when men gain what they toil for or know that they will use it.'"

[9] In a country such as the United States of America, with 50 subdivisions, it can enjoy over four dozen different "laboratories." This is the case for federalism, vis-à-vis centralism.

[10] In his text on ethics, Shafer-Landau (2015, p. 120) uses a specific concept for these values: Optimificality. This term means not only optimal, but that given all the shortcomings, it preserves some optimality without reducing others' rights or liberties.

[11] The notion that the 'people' choose the laws that govern them has been shown to be incoherent by Murray Rothbard (2009A). See also (Futerman and Block, 2021, p. 120) and (Rothbard, 2005, pp. 282 – 290).

[12] Often, even if an objective ethics or an optimal bundle of sticks exists, when it is unknown, and hard to discover, some laws, even if arbitrary, have to be established to avoid serious social problems. For instance, if statutory rape laws are justified at all, we know that if you go to bed with a 5-year-old girl, you are guilty of a violation thereof. If you do so with a 25-year-old woman, whatever you are, you are not a statutory rapist. But what is the proper cut-off point? Is it puberty at around 13? Is it 15, 17, 19? The point is that there is no exact age that may be deduced from any civilized principles. Also, some 16-year-olds are more mature than some 28-year-olds, so what is the real principle behind such laws? For more on this, see (Block and Barnett, 2008).

philosophers don't think is justified or efficient. This is to be contrasted with a copyright: In the latter, I agree through the purchase or use of the item to limit my use in specific ways that both buyer and seller agree on through contract (Rothbard, 2009B, pp. 745 – 754).[13]

Private property rights allow individuals the greatest freedom to act in accordance with their wills. It also allows a complex social system to operate with much less friction, as individuals can freely cooperate with whomever they choose on their own terms. Without it, many conflicts would be very hard, if not impossible, to resolve. This is not so surprising, as Aristotle notes (1992, pp. 114 – 115): "[T]he general principle should be that of private ownership. Responsibility for looking after property, if distributed over many individuals, will not lead to mutual recriminations; on the contrary, with every man busy with his own, there will be increased effort all around." Aristotle understood well that individuals care about the property they own and will be motivated to cultivate it and garner the most personal utility[14] from it (Evangeliou, 1995).

In effect, private property also offers the standard response to expropriation, aggression, and torts. It simplifies litigation, and each party can form a general expectation of what he owes others for damages and injuries caused by his property, or of the rules he should set in his own domain.

It further outlines human liberty: People are free to act on their private property, and on objects not currently owned by others, or to engage in title-transfer interactions. When these freedoms are in force, one needs only to show that he is the rightful owner of the property, and we are expected to cooperate, or at least not impede; it should not have to be further justified or approved of, and that lends it its power.

[13] The distinction between copyrights and trademarks or intellectual property will be discussed more rigorously in a forthcoming article.

[14] Aristotle talks about these perfect ultimate ends in the introduction to the Nicomachean Ethics (Aristotle, 2014, pp. 215 - 217). See also (Kenny, 1966; 1992).

### (2.4) The Economic Functions of Private Property.

So much for legal issues, we now consider some economic ones. The institution of private property is one of the central pillars of a free economy. It is necessary for any market guided by the price mechanism to emerge. Ludwig von Mises declares that (1998, p. 678) "[p]rivate ownership of the means of production is the fundamental institution of the market economy. It is the institution the presence of which characterizes the market economy as such." This is because a free market economy[15] is characterized by voluntary exchange between individuals and the functioning of the price system. In a market economy, individuals can transfer titles to one another only voluntarily if they own the objects of exchange. And the price system emerges from their subjective valuation of what they have and what they could trade for what others have. This is possible only when individuals are free to act on their property as they see fit and not merely respond to orders, thus giving rise to an economic price mechanism instead of the accounting management done in interventionist and socialist societies (Mises, 2009; 2012; Rothbard, 2011, pp. 815 - 825). Catallactics presupposes that all prices were reached without either state intervention or coercion; to the degree that the latter two are present, to that degree, the prices cannot be deemed to be market prices, and the exchanges are no longer catallactic.

[15] Henceforth, the market economy will mean a society in which all (or almost all) interpersonal interactions are voluntary. An unfree market will be considered a contradiction in terms.

Suppose that A owns[16] a set of goods $x(s) = \{x_1, x_2, \ldots\}(s)$, where 's' denotes all the circumstances at which his value scale obtains; and suppose that B owns another set of goods, $y(s') = \{y_1, y_2, \ldots\}(s')$.[17] It should be understood from this ordered set that in the circumstances *s*, A values $x_1$ over $x_2$, etc. In isolation, A or B may value their goods based on autistic exchange: What A sacrifices $x_n(s)$ for at *s* determines A's valuation of $x_n(s)$; valuing $x_n(s)$ over an arbitrary object, x' is a fleeting determination, and it merely tells us that A prefers x' over $x_n$ at s. Prices emerge when there is interpersonal exchange between A and B, over the goods they own. If A and B do not own, respectively, the sets x and y, then they cannot sacrifice their owned objects in exchange, and price allocation is arbitrary.

Without money, viz., when A and B engage in barter, there will be only prices based on the objects exchanged (the price of $x_i$ in terms of $y_j$, and vice versa), and these prices are determined at the moment of exchange. If A does not own $x_i$, and B does not own $y_j$, they can seek to own them and then exchange them. But if A and B are not able to own anything, they can only exchange in services, never in goods. Money facilitates their interpersonal exchange in both goods and services, since everything in the sets x and y can in principle have a monetary price, if anything else is valued higher than an arbitrary $x_i$ or $y_j$ in A or B's value scales.[18]

---

[16] For convenience, here, we will say that A owns, instead of controls, these objects.

[17] The list, x, is ordered through s: This means that $x(s_1) \neq x(s_2)$ if $s_1 \neq s_2$. The person, A, values $x_1$ over $x_2$ at $s_1$, but the situation might be reversed in $s_2$. It is impossible to limit the space of variable, *s*, to only some specific parameters like time, mood, etc. This is also not to say that *s* is an infinite space.

[18] A list is an unordered set. The set x can be ordered, but its ordering is arbitrary. However, the ordering of A and B's value scale is often taken to be from highest-valued goods or services to lowest. For a general outline of the Austrian theory of valuation and price-formation, see (Rothbard, 2009B, pp. 17 - 33, 106 - 137).

If, now, we are in such a private property system that the items in x cannot be appropriated (without incurring legal retribution) by anyone other than A, we may say that A not only controls, but legitimately owns, the items in x. In this way, private ownership does not merely make the exchange of goods possible; it makes interpersonal exchange efficient. When A and B each hold well-defined titles to their respective x(s) and y(s), every item (or a unit thereof) in these sets becomes a potential object of voluntary transfer, and money prices can emerge to reflect their changing marginal valuations. By contrast, where control over resources is communal or politically allocated, no single agent has the authority to alienate or pledge a given good, potential trades must pass through collective procedures or bureaucratic discretion, and the gains from trade are easily dissipated in delay, conflict, or indecision, or not be collected at all. Because private property concentrates disposal rights and residual gains and losses in its owners, it reduces bargaining and coordination frictions and channels resources to those who value them most. This prepares the ground for a further, closely related advantage of private property: its role in reducing social friction and enabling individuals to cooperate freely with one another, rather than constantly contesting control over assets.

### (2.5) Critique of some Textbook Definitions.

Not all economists start with private property in mind. Paul Samuelson and William Nordhaus define the science of economics as (2010, p. 4) "the study of how societies use scarce resources to produce valuable goods and services and *distribute them among different individuals*" [emphasis added]. It is all well and good to interpret "societies" as the collection of individual actors. But Samuelson and Nordhaus explicitly allow not only individuals, but also governments (and "society as a whole," via public goods) to own property. Yet ownership is an individual act of acquiring a scarce resource, which is now protected from others. When we say that governments own something, we implicitly mean that some of its officials, acting in the name of the state, have either confiscated a good (Rothbard, 2009A) or forbidden other citizens from owning that good (Alabbar, 2023). The state may then spend or act on this property, but only after such acts of (involuntary) confiscation or prohibition.

Moreover, this definition is insufficient for economics, as the method by which property is acquired and used differs fundamentally depending on ownership, whether public or private (Mises, 2015, pp. 37–55).

Samuelson and Nordhaus are interested in the government as a crucial acting agent (2010, pp. 36 – 38; p. 272; Ch. 16), and government "property" is certainly not private; nor is taxation, be it progressive or not, compatible with, to coin a term, *the sanctity of property rights*, as advocates of libertarianism hold. Their definition, which builds on Lionel Robbins', is deployed throughout the text to justify public spending, taxation, steady inflation, and similar interventions. These concepts are antithetical to private property (Hoppe, 1989).

In contrast, an example of an economic analysis that is based squarely on private property is Rothbard's.[19] On this view, a private

---

[19] Thus, Rothbard (2009B, pp. 91-92): "In a contractual society, each individual benefits by the exchange-contract that he makes. Each individual is an actor free to make his own decisions at every step of the way. Thus, the relations among people in an unhampered market are "symmetrical"; there is equality in the sense that each person has equal power to make his own exchange-decisions. This is in contrast to a hegemonic relationship, where power is asymmetrical—where the dictator makes all the decisions for his subjects except the one decision to obey, as it were, at bayonet point.

"Thus, the distinguishing features of the contractual society, of the unhampered market, are self-responsibility, freedom from violence, full power to make one's own decisions (except the decision property system is undermined to the extent that the state, or other individuals in a society, limit the owner's freedom to deal with his property in nonaggressive ways.

Returning to our concept of *the sanctity of private property*, a society holds this belief only insofar as its legal system favors no social good over the preservation of its individuals' private property.[20] Nozick (2013, pp. 33 - 34) calls this the libertarian constraint: that the ultimate decision made regarding an owned object is reserved to the individual owner, regardless of any other "social good" under this constraint.[21] If any

---

to institute violence against another), and benefits for all participating individuals. The distinguishing features of a hegemonic society are the rule of violence, the surrender of the power to make one's own decisions to a dictator, and exploitation of subjects for the benefit of the masters. It will be seen below that existing societies may be totally hegemonic, totally contractual, or various mixtures of different degrees of the two, and the nature and consequences of these various "mixed economies" and totally hegemonic societies will be analyzed.

"Before we examine the exchange process further, it must be considered that, in order for a person to exchange anything, he must first possess it, or own it. He gives up the ownership of good X in order to obtain the ownership of good Y. Ownership by one or more owners implies exclusive control and use of the goods owned, and the goods owned are known as property. Freedom from violence implies that no one may seize the property of another by means of violence or the threat of violence and that each person's property is safe, or "secure," from such aggression."

[20] The economists and philosophers of the Austro-Libertarian School have a distinctive conception of property, on which we draw. We can perhaps say that their thought exemplifies this 'sanctity of property' doctrine to a limit much wider than most (if not all) other social thinkers. However, this central position of property is not unique to the Austro-Libertarians (Rose, 1996), but as is well known, this school of thought makes the least allowances for the abrogation of property rights.

[21] To quote Waldron (2023): "In a system of private property, the person to whom a given object is assigned (e.g., the person who found it or made it) has control over the object: it is for her to decide what should be done with it. In exercising this authority, she is not understood to be acting as an agent or official of the society. She may act on her own initiative without giving anyone else an explanation, or she may enter into cooperative arrangements with others, just as she likes. She may even transfer this right of decision to someone else, in which case that person acquires the same rights she had. In general the right of a proprietor to decide as she pleases about the resource that she owns applies whether or not others are affected by her decision. If Jennifer owns a steel factory, it is for her to decide (in her own interest) whether to close it or to keep the plant operating, even though a decision to close may have the gravest impact on her employees and on the prosperity of the local community."

institution is allowed to override the owner's decisions on his property without suffering legal or retributive consequences, then property is undermined in such a society to the degree of violation.[22]

Furthermore, in one of the most widely used texts for undergraduate economics, private property is defined as "the right of private persons and firms to obtain, own, control, employ, dispose of, and bequeath land, capital, and *other property*" (McConnell, Brue, and Flynn, 2015, p. G13) [italics added].

The last clause—"and other property"—does most of the work: as with any concept, the definition of private property depends on what is allowed to fall under this residual category.[23] Drafting the definition in this way, so as to anticipate unspecified future contingencies, widens the definiendum and leaves it with an open texture (Waismann, 1945).[24] For "Finally, against the zeal of some reformers and some legal theorists who discuss the law of 'takings,' I shall argue that any overriding of an existing property right must either be with the right-holder's consent or else be accompanied by just compensation. In con-sequence, where just compensation is impossible, and consent cannot be obtained, no overriding of the right is justifiable."

---

[22] This rendition questions what some philosophers of law, like Lawrence Becker (2014, p. 2), have given:

Here the issue seems muddled: If the right holder consents to the 'takings,' it must be considered as donation or a trade for the just or market price of this specific inhomogeneous good, and strictly not 'takings.' Just compensation of a 'takings' is logically impossible: Without the right owner agreeing to any compensation, it is impossible to know the value of the good as esteemed by the right holder, and subsequently, no 'price' may emerge.

[23] Indeed, that sneaky clause makes the term 'property' even more vague.

[24] In law, openness makes the problem of forming a definition into a decision problem. Thus, it plays an important part for judges as they decide cases. Thus, Hart (2012) gives the example of how the definition of vehicle, as an open-texture, can develop on a case-by-case basis: "Faced with the question whether the rule prohibiting the use of vehicles in the park is applicable to some combination of circumstances in which it appears indeterminate, all that the person called upon to answer can do is to consider (as does one who makes use of a precedent) whether the present case resembles the plain case 'sufficiently' in 'relevant' respects. The discretion thus left to him by language may be very wide; so that if he applies the rule, the conclusion, even though it may not be arbitrary or irrational, is in effect a choice. He chooses to add to a line of cases a new case because of resemblances which can reasonably be defended as both legally relevant and sufficiently close. In the case of legal rules, the criteria of relevance and closeness of resemblance depend on many complex factors running through the legal system and on the aims or purpose

our purposes, however, the crucial point is that private property, understood as exclusive control over particular scarce resources, is ontologically distinct from the broader notion of "property rights," and marking this distinction is one of the central aims of this article.

A proper definition of property is essential to any economic and legal system, as it helps societies converge toward more efficient systems of allocation and adjudication. Economists and philosophers of the Austro-Libertarian School have a distinctive conception of property, on which we will draw. Our interest in their interconnected theories stems from their shared claim to an original, objective standard that aims to derive the essence of property from first principles rather than merely extending an inherited tradition of thought (Mises, 1998; Rothbard, 2005; 2009B). A further reason for focusing on them is their unusually vigorous defense of the institution of private property in contemporary economics and political philosophy. Once their respective contributions are brought together, a typical structure emerges. To foreshadow the forthcoming discussion, we will articulate this structure as a theory of property based on term-setting. At the same time, they leave underdeveloped an aspect we regard as essential, namely, the *openness* of property relations.

## 3. Philosophical and Ontological Aspects of Property.

Before we may proceed further, let us first discuss two senses of the term 'property,' since it is often used to mean two distinct things. We generally speak of *property* in two senses: An attributive sense, indicating ownership (*e.g.*, this car is *mine* or *yours*), and a descriptive sense,

---

which may be attributed to the rule. To characterize these would be to characterize whatever is specific in or peculiar to legal reasoning."

indicating a quality (*e.g.*, this car is *red* or *big*). In this article, our concern will be with the positive and normative aspects of the former sense: i.e., how our species exhibits (or normatively, should exhibit) this feature.[25] Property, in the latter sense, is metaphysical and is almost exclusively used by philosophers. We choose to refer to these descriptive properties as qualities to avoid confusion. The use of qualities will be essential to our definition of property.

In the previous section, we saw excerpts discussing property *rights* and their importance. But what about property, per se? A definition is needed in order to know what precisely is being referred to when speaking about it.

And here lies the problem (Bell and Parchomovsky, 2005): "property law often seems to suffer from a characteristic disease of legal categories; everyone knows what it is, but no one can define it."[26] But now we have to ask: How can anyone *know* what property is if no one can define it?

Philosophers and mathematicians have a way to deal with similar problems: Something can be understood intuitively without being properly definable. After all, the geometrical *point* is undefinable,[27] but we may derive a definition of the geometrical line from our intuitive understanding of a point. The same goes for sets in set theory, and many other fundamental objects of the logical sciences. Mathematical objects are often defined by their qualities, or by the conditions they satisfy. Philosophers often discuss ideas that lose some degree of vagueness after analysis. Knowledge is one such interesting example: Previously defined as *justified true belief*, Gettier (1963) found beliefs that satisfy the definition without being counted as knowledge, because they are mistaken, which encouraged some epistemologists to refine the definition (Plantinga, 1993A; 1993B).

---

[25] Metaphysicians study these descriptive properties of concrete and abstract objects, and natural scientists study the objective qualities of material objects and the universal laws that govern them. In our exploration, we will not be interested in what the physicist or the metaphysician study. The things that make an object *rightly* belong to someone are not inherent, nor necessarily essential to the object itself.

[26] No one can fully satisfactorily define pornography, but most people claim they can tell it when they see it, to paraphrase Justice Potter Stewart. The devil, however, is in the details: As most intuitive definitions of property are vague enough for the associated rights to be abridged.

[27] Indeed, such definitions are often circular, push the problem further backwards, or relegate it into human intuition. For example, Euclid defines the geometric point as 'that which has no dimension.' But then, what does dimension mean? We encounter many things which we call points, and then leave its understanding to intuition.

In like manner, we can make our intuitions of property less and less fuzzy to refine our corollary definitions. But care should be taken: A study of what things we typically call property is helpful to a degree until we exceed that limit and include more objects in the category than is needed, or overfit the definition and not allow certain things in the future to be appropriated that are of a nature we can not conceive of today. This is, after all, the point of a good definition: It offers the criteria that group together all and only the relevant cases.

This leads us to a related linguistic and epistemological problem. Wittgenstein (1991, §§65–70) considers the multifaceted phenomena we call games and argues that they are connected by overlapping "family resemblances" rather than by a single common core that would ground a unified definition. This diagnosis remains highly relevant: it shows that we should not expect a simple essence shared by all games. Yet Rowe (1992) argues that, even granting Wittgenstein's point, we can still group these cases in a principled way and extract a satisfactory definition of "game."

Morris Weitz (1956) applies Wittgenstein's point to art, asking rhetorically whether there is any empirical question that could decisively determine that something is a work of art.[28] He argues that art is an *open concept*: we face a choice between fixing a definition that would inevitably constrain innovation and allowing the concept's criteria to evolve alongside artistic practice; i.e., settling borderline cases is less a discovery of the essences of artifacts than a practical decision about which criteria we choose to call artistic.

Peter Milonni (2004) resolves an analogous definitional ambiguity by decomposition. Instead of treating "the speed of light" as a monolith admitting one definition, he distinguishes several physically different velocities in media, each with its own definition and equation. On this view, many disputes over "the" speed of light reduce to a verbal compression: different experiments measure different velocities, but their results are reported under the same name.[29]

This offers us a recourse for our original problem. Often, people differ over what is beautiful and what isn't, and whether beauty is an objective quality. It is easier to converge on what constitutes physical matter and where any object's physical form lies, except in very exotic cases (e.g. extreme cases of relativistic velocities and in the microscopic limit, to which we relegate the discussion to future generations).

By analogy, one might suspect that "property" is likewise resistant to definition. It is undoubtedly challenging to frame a definition that covers all things that may be called property; indeed, an entirely "atomic" definition that introduces no undefined concepts at all is impossible, as noted above. This, however, is a general feature of conceptual analysis and need not detain us.

---

[28] For a more extended discussion, see (Adajian, 2022), and the reference and quotation in footnote 22.

[29] The velocities delineated by Milonni are: Vacuum, phase, group, wavefront, energy-transport, and signal velocities. All but the vacuum and signal velocities may exceed the set value, $c_0$.

Although the literature contains many refinements and sub-variants of the ontological method alluded to in (§1), at a higher level of abstraction, we can distinguish three main approaches: The legal-positivist approach (a) treats “property” as whatever the prevailing legal order designates as such, by granting its owner the associated rights; the functional approach (b) understands property in terms of the roles it plays in social life, for example by asking which holdings realize the six social functions identified in (§2.1); and, the procedural approach (c) conceives property in terms of its origin in specified schemes of original acquisition and title-transfer.

Having drawn this tripartite map, we can now narrow our focus. The legal-positivist approach (a) is, for present purposes, too thin: a legal order can, in principle, attach the label “property” and a bundle of associated rights to almost anything, and can modify or withdraw those rights at will. This makes (a) valuable as a descriptive account of how particular legal systems actually speak. Still, it yields no substantive constraint on what ought to count as property, except for those wedded to Étatist predilections.

The Benthamite implication is problematic in three respects. First, it sits uneasily with the earlier concept of property’s ‘sanctity,’ since it recasts ownership as a revocable set of permissions contingent on continuing legislative will, rather than as a stable domain of authority that law is meant to recognize and secure. Second, it undermines the central social point of property—namely, the liberty and security it affords by stabilizing expectations (as captured by the six functions discussed above): if people come to regard their holdings as merely provisional grants, subject to discretionary alteration, the resulting insecurity invites conflict rather than coordination, with a potential decivilizing tendency.[30] Third, it misdescribes the historical development of property: at least in the common-law tradition, legal doctrine typically crystallizes and systematizes practices and expectations that are already operative in social life, rather than creating the institution ex nihilo (Hayek, 2011, p. 115; 1998, pp. 85 - 88). This leads us to reject the legitimacy of this approach altogether.

A more robust candidate is the functional approach (b), on which "property" denotes those holdings that in fact realize its functional roles. Yet this functional characterization is not independent of the procedural approach (c). Once a class of holdings has been picked out by a justified scheme of original acquisition and transfer, those holdings will, ceteris paribus, tend to perform the social tasks we associate with private property; if they did not, the scheme would be defective as a property-acquisition procedure. In this sense, the functional approach presupposes the procedural one: it describes what properly acquired holdings do in a social order. For definitional and normative purposes, then, the procedural approach is conceptually prior and sufficient by itself. The remaining question is which acquisition and transfer procedure should we regard as justified?

We can draw our first intuitions from John Locke's famous passage on the acquisition of property (Locke, 2014, p. 731): "Though the earth and all inferior creatures be common to all men, yet every man has a property in his own person: this nobody has any right to but himself. The labour of his body, and the work of his hands, we may say, are properly his. Whatsoever then he removes out of the state that Nature hath provided, and left it in, he hath mixed his labour with it, and joined to it something that is his own, and thereby makes it his property. It being by him removed from the common state Nature placed it in, it hath by this labour something annexed to it that excludes the common right of other men: for this 'labour' being the unquestionable property of the labourer, no man but he can have a right to what that is once joined to, at least where there is enough, and as good, left in common for others."

[30] The notion of decivilization is borrowed from John Keegan (1993), where it is defined to be the tendency to resolve conflicts through violence, instead of negotiations and mediating institutions.

Although Locke's passage[31] can help us, and it is often a starting point for modern property theorists, it is not entirely coherent by itself. How can we define property as something that is owned once our *labor* has been *mixed* with it, if labor itself is considered property? This is circular to an advanced degree,[32] because, in Locke's formulation, the justification of homesteading is predicated on the fact that our labor is property and that we own it. Do we first mix our labor with our labor before acquiring it as property? If yes, we have no account of when that happens; if no, then there are procedures that precede this, and they become the central schemes. Furthermore, Waldron demonstrates that the idea of mixing one's labor as a form of acquisition is fundamentally confused (Waldron, 1983); he also points out that the labor-mixing axiom cannot be quantified, which makes it useless. We still hesitate to abandon this term since it forms our natural conception.

The naïve (procedural) notion of private property we can initially extract may be thus summarized: Private property is any object that is produced—i.e., *sufficiently* used by a human being to satisfy an end, or transformed into a form it is useful in, cared for so that it might be used in the future—or received in title-transfer, provided that this object was not previously subjected to such use (production) by an unconsenting third party. This is quite a mouthful, and its terms and clauses need much explanation and analysis. It is to be understood that different objects require different investments in usage before they are owned. A land that is crossed is not owned by the action of crossing, but one that is seeded and tilled is. Further, the addition of 'human beings' to the definition is arbitrary. That the use serves an end is a redundancy. (Can we conceive of any action that is not a means to an end?) However, when we talk about property, we mean these actions, and the labor-mixing is an abbreviated code for them.

[31] The Talmudic tractate "Baba Metzia" is also apropos.
[32] For more on definition circularity, see (Gupta, 2023).

Furthermore, as Ludwig von Mises notes, it is exceedingly difficult to trace the history of any given object far enough into the past to determine whether it was never previously appropriated or whether it passed through some act of violent expropriation (2015, pp. 42 – 43): "All ownership derives from occupation and violence. When we consider the natural components of goods, apart from the labour components they contain, and when we follow the legal title back, we must necessarily arrive at a point where this title originated in the appropriation of goods accessible to all. Before that we may encounter a forcible expropriation from a predecessor whose ownership we can in its turn trace to earlier appropriation or robbery." There has to be a specific cut-off to the final clause of the naïve definition, since much legitimate property today has been violently expropriated. It is obvious, then, that our use of the term property has a certain degree of pragmatism: Private property as an institution would be utterly futile if ownership required that we spend infinite time and resources (which is an impossibility) to locate all previous or legitimate owners to make sure there was no aggression at any point in time.[33]

[33] The market has already solved this problem with title-insurances. See (Block, 2009, pp. 412-413).

It is not as easy, however, to appropriate previously unowned objects from nature nowadays.[34] So, we have to consider property as it is commonly seen today. Indeed, it is much easier to consider the kinds of things we may call property in our modern industrial age. So let us start at the present, disregarding the illegitimacy of objects owned through aggression, as in Mises' quotation above.

Most of our private property today is acquired by title-transfer: e.g., inheritance, gifts, private benefactions, government grants and subsidies, and trade. Let us consider the last: Take purchased items (e.g., a book, some fruit, or a laptop): This title-transfer through an exchange might well be the quintessential form of acquisition today. We can easily see the transferred goods as the purchaser's property, with the purchaser having complete ownership of the items. The current owner will have so much liberty with them. He can eat an orange, squeeze it for juice, contemplate it, or throw it away. These liberties are accessible to most people worldwide.

But as we have noted, ownership is constrained by the conditions imposed on it. Is the purchaser completely free to do whatever he wants with the orange? He has so many liberties, and most citizens have a duty not to intervene in his usage, that we are eager to say yes, so long as he does not *harm*[35] another or limit the freedoms of his peers. Consider this question, then: Is he free to get juice from the orange and sell it at a stand? He is not at liberty to do so in many places in the world due to modern regulations.[36] Does that make the setters of the rules for how such items may be used, in effect, partial owners, reducing the degree to which the former owns said items? They are, after all, the ultimate decision-makers here regarding at least some of the possible usages. We arrive at the same conclusion as Rothbard (2009B, pp. 1276 - 1279) in saying that, inasmuch as an entity limits the use of a resource, it partially owns it. We will call this Harper's Principle: "What sense is there to calling it ownership if you cannot sell it, […] The corollary of the right of ownership is the right of disownership. So if I cannot sell a thing, it is evident that I do not really own it."[37]

[34] The glaring counterpoint libertarians would use is the *land as standing room*, however, we discard these objects as pseudo-property in a forthcoming article.

[35] Scare quotes around this word since we are entitled to harm each other in many ways. For example, in the marriage market: A and B both want to marry C. She can only marry one of them. The other is harmed. Ditto for competition in commerce. Customer X chooses between store Y and Z; the one not chosen is harmed. What we are not allowed to do in the free society is violate other people's rights; it is legally licit to harm them.

This brings us back to our concept of the sanctity (or presumed inviolability) of property within the Lockean framework. On that view, coercive restrictions are justified only where they prevent or remedy rights-violations, or where parties have voluntarily altered their rights through agreement.[38] The mere existence of a contract does not by itself warrant third-party coercion; rather, coercion is justified only when a party has consented to specific transfers, penalties, or enforcement terms, and then refuses to perform—so that the resulting interference aims to vindicate the agreed allocation of rights rather than to initiate force. By contrast, it is not clear why the alleged "aggression" in the lemonade-stand example would warrant government intervention, especially if the appeal is to a purported social contract that, on this account, lacks binding

[36] In the state of Kuwait, for example, it is impermissible to sell fruits without a license: https://www.baladia.gov.kw/sites/ar/municipalityServices/Pages/municipalityRegulations/pdfs/food.pdf (Retrieved 22nd of July, 2025.) This was done, allegedly, to combat the sales of food items (mostly fruit stands for watermelons, sunflower seeds, and similar snacks) by the *Bidoon* (Alabbar, 2022A; 2022B). Similar cases apply in many states in the US. See, for example, (Fleming, 1974). The example demonstrates that individuals are not at liberty to practice many acts that are *not* per se aggressive.

[37] From (Harper, 1949, p. 106); quoted from (Rothbard, 2009B, p. 1278).

[38] For a reinterpretation of the 'enough and as good' clause in Locke's Homesteading principle, see (Waldron, 1979)

consent. Yet, as we have noted, regulations of this sort abound across myriad other industries.

Many legal theorists and political philosophers, however, answer in the negative when asked if these regulations make the government a party to the ownership of the regulated objects: Although private property should be immune from arbitrary control and expropriation, the liberties associated with ownership must be in accordance with the law. This answer is, for example, unsatisfactory to a libertarian who believes that the laws should not conflict with human liberties, as it matters very little whether the controller or expropriator is a king, a judge, a bureaucrat, a member of parliament, or an otherwise normal citizen. Furthermore, the owner has to deal with more than just common law: He also has to deal with legislation and civil law, and that goes to show how varying property rights are around the world.

Indeed, the term "property" was not properly defined in English law until the late Seventeenth century, and even then it remained elusive, as it involved the crown in ownership and allowed it to expropriate "private" property (Aylmer, 1980). The definition of property is often linked to the private *rights* of individuals (and, erroneously, of firms) or it extends to the objects owned by a community or the government, as in the legal-positivist approach. It needs to be said that without fully knowing what set of objects may be acquired as property, the institution of private property still flourishes since we appreciate the rights of individuals with respect to the objects associated with them, albeit to a lesser degree.[39] However, without a proper definition, some long-lasting problems persist. Nevertheless, the common-law courts allowed property rights to be defined as floating, and the rights would become ever more precise and more explicit whenever a court trial or legislation warrants it.

[39] An analogy: We have survived for tens of thousands of years without obstetricians, but having them helps us reduce many mortalities during pregnancy or childbirth.

# 4. Necessary Conditions of the Lockean Scheme.

## (4.1) Openness and Vagueness.

Finding the set of individually necessary and jointly sufficient conditions that would make an object *private property* and its owner the sole proprietor[40] would be useful to jurists and legal theorists, as it would help us properly categorize human rights. Some libertarians, for example, ground all human rights under property rights (noting that they consider the body of an individual to be his own private property (Hoppe, 2006; Rothbard, 2002)). Furthermore, it would also be useful in ruling out some claimed rights and instituting others.

However, it is not obvious whether this question has a definitive answer. The fact that some concepts evolve over time, as technology, culture, and traditions change, makes them open-textured. We will call this feature *openness.* We again remind the reader that the focus will be on a specific procedural approach, namely the classical scheme of acquisition, viz., the Lockean Homesteading Scheme.

In (§1), we outlined three methods of defining property: Utilitarian, Just, and Ontological. From the ontological method, we extracted in (§3) three main approaches: Legal-Positivist, Functional, and Procedural. We rejected the first therein and showed that the third is analytically prior (and subsumes) the second.

---

[40] i.e., to prove that private property is a pellucidly clear concept.

Many scholars have addressed the vagueness of the term 'property' and other legal concepts in the philosophy of law (Penner, 2000, Ch. 1; Hart, 2012, Ch. 7). Vagueness differs from openness in that a concept is vague when more information is required in order to increase the precision regarding the essence of the concept, whereas it is an open-texture when the boundaries of the concept becomes a decision problem, rather than an epistemological one (Waismann, 1945; Shapiro and Roberts, 2019; Weitz, 1956). We will return to vagueness in the next subsections as we revisit the definition of property. Now, we will limit the openness of property by observing some necessary conditions from Locke's (2014) conception.

Property has both objective and subjective qualities. Prices, for example, emerge from the subjective valuation of owned objects. (The concept of objective value has long been redundant, except explicitly by those who champion the labor theory of value and implicitly by those who follow the doctrine of fixed costs.) Some objective qualities are relevant, and they inform the Lockean theory, as will be demonstrated below.

### (4.2) Three Ontological Conditions.

We have so far considered property to be something physical; property *rights*, of course, being abstract, are not property. The procedural approach, or Locke's homesteading scheme, specifically forces us to consider first *use*[41] in acquisition. This relegates the vagueness to *use*, as the example above shows in the definitions of the point in mathematics and knowledge in epistemology, or creates an ambiguity as in the definition of the speed of light in physics. Although we may be tempted to adopt their solutions by further analyzing the concept or showing that there may be many definitions of property, we will address the term 'use' head-on.

---

[41] Or mixing one's labor with the object.

Ontologists separate the world into categories, among which are the concretes (objects that exist in space and time) and the abstracts (existents that are not concrete): A pen is a concrete object, whereas qualities, relations, numbers, and structures are all abstract categories, containing infinitely many objects (Grossmann, 1992).

To *use* something at all in the Lockean sense is to physically interact with an object, and thus, that thing necessarily has to be concrete. Exploiting a relation (say, between a customer and his desire for a consumer good) has to be ruled out from the action of 'use,' or mixing one's labor, since one cannot mix one's labor with an abstract object.[42] As concrete objects, we interact only with concrete objects, which have certain qualities, specific relations, etc, and therefore, only these concrete objects can be acquired. By definition, it is impossible to use abstracta, as these objects cannot change spatiotemporally.[43]

So, the first necessary condition is *concreteness*; i.e., that such an object exists, and is concrete. This condition is by no means self-evident: Something indeed has to exist for it to be property, but it is not apparent that the object has to be concrete. In fact, concreteness is a controversial condition, and intellectual property or real property laws, for example, are a testament to that.[44] Since only concrete objects can be used, only concrete objects can be owned. This form of ownership, through first use, has been historically called *homesteading*.

---

[42] Mixing one's labor with an object is a roundabout way of saying transforming an object, or using it in production.

[43] What about abstract objects that exist in time, but not in space? Grossmann (1992) gives the example of a thought. An idea, for example, does not exist spatiotemporally, but a thought is conceived at a certain time (when it is thought) and vanishes once the thought is no longer contemplated. The qualities of these thoughts do not exist spatiotemporally, but the thought exists in time (not in space, however, as thoughts cannot be localized). We must rule these out, since it is inconceivable to us how one may use a thought in vacuo. (An Aside: There are no converse objects — objects that exist in space but not in time.)

It may be seen now that concreteness is not sufficient by itself. In a sense, concreteness means that something can, in principle, be homesteaded. In a wider sense, concreteness can also be interpreted as practical homesteadability. But there are other practical concerns, the most important of which follows.

Since objects can always be grouped or divided, another condition becomes apparent. For example, if one dips a bucket into a river, he has used the river, but he only owns the water in the bucket; and by feeding some cattle from a wild flock, one owns the fed or tended cattle, but not necessarily the whole flock.

This leads us to the second condition: *individuation*. The *Principium Individuationis* allows us to separate objects into precisely identifiable quantities that are spatiotemporally distinct from one another. We may loosely say that acquisition individuates objects.[45] Individuation may also be loosely considered the ability to set terms for a part of a continuous object or a collection of separable objects.

The third condition is formally ontological, since it follows from the definition: *Res Nullius*, or *unownedness*. Assuming that something is to be acquired through first use, it ought to be unowned. Of course, first use here can also mean that the object was previously owned and then relinquished, or relinquished through exchange or title transfer. An object cannot be homesteaded if it is currently owned.

[44] Or consider the case of fiat money: By owning a $1 bill, some people think that they are entitled to the purchasing power of that dollar; while the physical dollar may be exchanged with other items in the market, one cannot own any value or purchasing power beside the physical dollar itself.

[45] Rothbard (2002) gives another example on individuation by considering what we call in a forthcoming article 'Columbian ownership,' wherein someone may own a whole continent by setting foot on its shore and homesteading a smaller piece of land.

### (4.3) Praxeological and Ethical Conditions.

To narrow the scope of this article, we will briefly mention two final necessary conditions that warrant separate attention but do not follow from Locke's formulation. One relates to motive and the other to stability.

The fourth condition is what motivates individuals to own an object, without which, no object will be owned: Value. Objects are subjectively valued iff they are scarce, and they can be placed at some point in a production plan — either at its end as a higher-order good, or at its beginning as a lower-order good. As for scarcity, humans have no use appropriating things that are of 'unlimited' abundance, and by the act of possessing them, we use scarce resources (even if time alone) to transform them into scarce items.[46] Production and scarcity are intimately linked by the same mechanism.[47]

---

[46] Air, for example, is not a scarce resource. Rothbard (2009B) shows that humans count objects that are *infinitely* abundant as a general condition of human welfare. (By infinite, this is a subjective estimation, and as soon as this estimation no longer obtains, air ceases to be abundant.) By filling a balloon with air, the collection of molecules within the balloon, which used to be part of the abundant air, became scarce. The appropriation and possession of parts of an abundant object necessarily make the appropriated bit scarce, by the principium individuationis.

[47] Production, in the Austrian context, simply refers to moving the object one step higher in the ladder of goods. However, scarcity is intimately related to the process of production, due to the law of diminishing returns and the principle of opportunity cost. In another vein, consider this: We do not (and cannot) own *infinitely* abundant resources like air or sunlight (Rothbard, 2009B, pp. 1277): "Ownership is the ultimate control and direction of a resource. The owner of a property is its ultimate director, regardless of legal fictions to the contrary. In the purely free society, resources so abundant as to serve as general conditions of human welfare would remain unowned. Scarce resources, on the other hand, would be owned on the following principles: self-ownership of each person by himself; self-ownership of a person's created or transformed property; first ownership of previously unowned land by its first user or transformer."

The fifth (ethical) condition is the one responsible for the continuation of the enterprise, and therefore, we call it the stability condition. It is usually called the *excludability* quality; however, more generally, it is the owner's ability to set terms that are socially or legally binding, to organize the use of the object. This final condition allows the enterprise to continue by giving owners legal and social power to exercise their rights on their owned objects. This condition is far-reaching, as it forbids social and legal intervention in the exercise of one's freedoms toward the objects he owns, regardless of the ends sought by the intervention.

## 5. Conclusion: The Definition of Property, Revisited.

We may now sum up what we have done. An object is said to be the *private property* of the proprietor if he homesteads it or if the title was transferred to him by someone else who had homesteaded it. For an object to be homesteadable, it has to be concrete, individuated, unowned, (subjectively) valued, and controllable. This simple act of (first) using the object reallocates it to particular individuals, who may exclude others from using these objects or set terms for their use. The six functions of property concern the optimization of production, which is the ultimate goal of property.

In the introduction (§1), we asked three main questions: What things may be appropriated? What actions allow us to appropriate an object? And who may control the object appropriated? We used ontology to answer the first questions, and offered hints of a solution to the second and third questions, which are to be answered in upcoming articles. Then, we showed the most important reasons why scholars are in dispute regarding these questions and why a universal definition (i.e., accepted by all) is very hard to obtain:  People differ in ethics; property rights form systems of power; and for historical reasons that we may call collective intellectual inertia. Finally, we separated definitions of property based on efficiency, justice, and those due to procedure, which were our ultimate concern here.

In order to understand why property is such an important institution, we listed the six main social functions of property in (§2.1): The allocation of control, the creation of an incentive structure, conflict resolution, the facilitation of economic calculation, the ability to set rules, and the freedom of association. These functions offer the highest degree of protection and predictability that the institution of property can afford.

Then, in (§2.2), we demonstrated why property is necessarily a *social* institution, and that it is an evolved, and not an invented, tradition; (§2.3) that property looks similar across legal systems, but that it differs in some *sticks* in the bundle, and that the implications of these small deviations can become large in any complex system; (§2.4) how property is essential to an economy in that it allows people to subjectively value objects through which the price structure emerges, and that without it, exchange can only be in terms of services, which is much less efficient; we concluded the second section in (§2.5) where we looked at two textbook definitions, and how they contain concepts that are antithetical to the pure form of property: The first is that some definitions do not elevate property to its position as a central institution (which we have called the sanctity of property), defined as the degree to which societies have the disposition to protect and honor each other's property rights, and the second is the degree of openness associated with defining property.

Definitions can be found by analysis, construction, or decision. The main purpose of (§3) was to consider these types of definitions. We focused on the analytical method, which required us to study the ontological aspects of property. In it, we saw the legal positivist, functional, and procedural approaches, and showed that the latter is analytically prior and sufficient by itself. However, there are fatal flaws with Locke's Homesteading scheme, which we had to address.

Property is often used as a paradigm for all rights. In (§4), we considered the objective and subjective qualities of property. Then, we analyzed Locke's homesteading scheme by studying how labor mixing (which we call *use*) is performed. By understanding what *use* means, we immediately eliminate abstracta from appropriation. Then, we listed five necessary conditions for an object to be owned.

In the end, we can say that in the Lockean scheme, private property is an object with the aforementioned subjective and objective qualities that grants the owner the ability to set the terms under which these resources may be used and satisfies these five necessary conditions.[48]

**Declaration of Competing Interests: The authors declare none.**

**Declaration of Funding: The authors declare none.**

**Author Contributions: All authors contributed to each section of the paper.**

## Acknowledgments.

We are grateful to the editors for their helpful suggestion to divide what was originally a single, longer manuscript into a series of articles. The present article, together with the four that will follow, emerged from that earlier unpublished work. The authors would like to thank Abdullah Al-Shimmery and Dr. Mohammad Al-Khaldi for their comments and criticism of an earlier version of this paper. Needless to say, but we will say it anyway, the present authors, alone, are responsible for all remaining errors and infelicities.

[48] This article branched out from a longer unpublished manuscript. The article was getting too long, and so, we segmented it into five separate articles. Following this introductory article, we will consider some immediate consequences in the next one, wherein we separate objects said to be property into real-, quasi-, and pseudo-property, following the property rights paradigm. Then, in the one after, we properly define 'use,' by considering methodological singularism, and actions of acquisition. After that, we discuss collective ownership through methodological individualism. In the final article, we will study the ethical dimensions of private property.

# References.

**Adajian, Thomas**. 2022. The Definition of Art, Stanford Encyclopedia of Philosophy.
**Alabbar, Adnan**. 2022A. The Private Lives of the Bidoon, Platform Post.
**Alabbar, Adnan**. 2022B. Nationality and Statelessness: The Kuwaiti Bidoon, Mises Wire.
**Alabbar, Adnan**. 2023. Much of the World's Oil Is Owned by Governments. There's No Good Reason for This, Mises Wire.
**Alchian, Armen A**. 1965. "Some Economics of Property Rights," Il Politico , DICEMBRE 1965, Vol. 30, No. 4, pp. 816-829.
**Alchian, Armen A., and Harold Demsetz**. 1973. "The Property Right Paradigm." The Journal of Economic History 33 (1): 16–27.
**Alexander, Gregory S**. 2009. "The Social-Obligation Norm in American Property Law." *Cornell Law Review* 94 (4): 745–819.
**Aristotle**. 1992. The Politics, translated by T. A. Sinclair, Penguin Books.
**Aristotle**. 2014. Aristotle's Ethics, Writings from the Complete Works, revised, edited, and with an introduction by Jonathan Barnes and Anthony Kenny, Princeton University Press.
**Aylmer, G. E**. 1980. The Meaning and Definition of "Property" in Seventeenth-Century England, Past & Present, Feb., 1980, No. 86 (Feb., 1980), pp. 87 – 97.
**Becker, Lawrence C**. 2014. Property Rights: Philosophic Foundations, Routledge.
**Bell, Abraham, and Gideon Parchomovsky**. 2005. A Theory of Property, 90 Cornell L. Rev. 531 (2005).
**Bentham, Jeremy.** 1838. The Works of Jeremy Bentham, published under the Superintendence of his Executor, John Bowring (Edinburgh: William Tait, 1838-1843). 11 vols. Vol. 1.
**Block, Walter E**. 1977. "Coase and Demsetz on Private Property Rights," The Journal of Libertarian Studies: An Interdisciplinary Review, Vol. I, No. 2, 1977, pp. 111-115.
**Block, Walter E**. 1995. "Ethics, Efficiency, Coasean Property Rights and Psychic Income: A Reply to Demsetz," Review of Austrian Economics, 8 (2): 61-125.
**Block, Walter E**. 2000. "Private Property Rights, Erroneous Interpretations, Morality and Economics: Reply to Demsetz," Quarterly Journal of Austrian Economics, Vol. 3, No. 1, Spring, pp. 63-78.
**Block, Walter E., and William Barnett II**. 2008. Continuums, Etica & Politica / Ethics & Politics, X, 2008, 1, pp. 151-166.
**Block, Walter E**. 2009. The Privatization of Roads and Highways, Ludwig von Mises Institute.
**Block, Walter E**. 2010. "A Response to Brooks' Support of Demsetz on the Coase Theorem." Dialogue, Vol. 2.
**Block, Walter E., and Peter Lothian Nelson**. 2015. Water Capitalism: The Case for Privatizing Oceans, Rivers, Lakes, and Aquifers, Lexington Books.
**Block, Walter E**. 2019. Property Rights: The Argument for Privatization, Palgrave Macmillan.
**Block, Walter E**. 2022A. Thick and Thin Libertarianism and the Non-Aggression Principle, Etica & Politica, XXIV, 2022, 2, pp. 31-54 ISSN: 1825-5167.
**Brooks, Michael**. 2007. Toward a Clarification of the Block-Demsetz Debate on Psychic Income and Externalities, Quarterly Journal of Austrian Economics, Vol. 10, pp. 223-233.
**Buckle, Stephen**. 1993. Natural Law and the Theory of Property: Grotius to Hume, Clarendon Press.
**Currier, Richard L**. 2017. Unbound: How Eight Technologies Made Us Human, Transformed Society, and Brought Our World to the Brink, Arcade.
**Demsetz, Harold**. 1967. "Toward a Theory of Property Rights." *The American Economic Review* 57 (2): 347–359.
**Demsetz, Harold**. 1979. Ethics and Efficiency in Property Rights Systems, in Time, Uncertainty, and Disequilibrium: Exploration of Austrian Themes, Mario J. Rizzo (ed.), Lexington Books.

**Demsetz, Harold**. 1997. “Block's Erroneous Interpretations,” Review of Austrian Economics, Vol. 10, No. 2, pp. 101-109.
**Evangeliou, Christos**. 1995. Even Friends Cannot Have All Things In Common: Aristotle’s Critique of Plato’s Republic, Presented to the SAGP meeting with the American Philological Association, San Diego, December 28, 1995.
**Fleming, Harold M**. 1972. Ten Thousand Commandments: A Story of the Antitrust Laws, Arno Press.
**Foss, Nicholai J., Peter G. Klein, Lasse B. Lien, Thomas Zellweger, and Todd Zenger**. 2020. Ownership Competence, Strat Mgmt J. 2021; 42: 302–328. https://doi.org/10.1002/smj.3222
**Futerman, Alan G., and Walter E. Block**. 2021. The Austro-Libertarian Point of View: Essays on Austrian Economics and Libertarianism, Springer.
**Gettier, Edmund L**. 1963. Is Justified True Belief Knowledge?, Analysis, Vol. 23, No. 6 (Jun., 1963), pp. 121-123.
**Grossmann, Reinhardt**. 1992. The Existence of the World: An Introduction to Ontology, Routledge.
**Gupta, Anil, and Stephen Mackereth**. 2023. "Definitions", The Stanford Encyclopedia of Philosophy (Fall 2023 Edition), Edward N. Zalta & Uri Nodelman (eds.).
**Hardin, Garrett**. 1968. “The Tragedy of the Commons.” Science 162 (1968): 1243–1248.
**Harper, F. A**. 1949. Liberty, a Path to Its Recovery, Irvington-on-Hudson, N.Y.: Foundation for Economic Education, 1949, reprinted by the Ludwig von Mises Institute, 2007.
**Harris, Cheryl I**. 1993. “Whiteness as Property.” *Harvard Law Review* 106 (8): 1707–1791.
**Hart, H. L. A**. 2012. The Concept of Law, 3rd ed., Clarendon Law Series, Oxford University Press.
**Hayek, Friedrich**. 1945. The Use of Knowledge in Society, The American Economic Review, Sep., 1945, Vol. 35, No. 4 (Sep., 1945), pp. 519- 530.
**Hayek, Friedrich**. 1988. The Fatal Conceit: The Errors of Socialism, University of Chicago Press.
**Hayek, Friedrich**. 1998. Law, Legislation, and Liberty: Volume 1 - Rules and Order, Routledge.
**Hayek, Friedrich**. 2011. The Constitution of Liberty, from The Collected Works of F. A. Hayek, Volume XVII, Ronald Hamowy (Ed.), University of Chicago Press.
**Hohfeld, Wesley Newcomb**. 1913. Some Fundamental Legal Conceptions as Applied in Judicial Reasoning, The Yale Law Journal, Vol. 23, No. 1 (Nov., 1913), pp. 16-59 (44 pages). https://doi.org/10.2307/785533
**Honoré, A. M**. 1961. “Ownership.” In Oxford Essays in Jurisprudence, edited by A. G. Guest. Oxford: Oxford University Press.
**Hoppe, Hans-Hermann.** 1989. Fallacies of the Public Goods Theory and the Production of Security, The Journal of Libertarian Studies, Vol. IX. No. 1 (Winter 1989)
**Hoppe, Hans-Hermann**. 2006. The Economics and Ethics of Private Property, 2nd edition, Ludwig von Mises Institute.
**Hoppe, Hans-Hermann**. 2009. The Private Production of Defense, Ludwig Von Mises Institute.
**Hoppe, Hans-Hermann**. 2015. A Short History of Man: Progress and Decline, Ludwig Von Mises Institute.
**Keegan, John**. 1993. A History of Warfare, Vintage Books.
**Kenny, Anthony J. P**. 1966. Happiness, Proceedings of the Aristotelian Society, New Series, Vol. 66 (1965 - 1966), pp. 93-102.
**Kenny, Anthony J. P**. 1992. Aristotle on the Perfect Life, Clarendon Press.
**Kinsella, N. Stephan**. 2001. Against Intellectual Property, Journal of Libertarian Studies, Volume 15, No. 2 (Spring 2001): 1–53.
**Landes, William M. and Richard A. Posner**. 2003. The Economic Structure of Intellectual Property Law, Belknap Press.
**Locke, John**. 2014. An Essay Concerning Human Understanding with the Second Treatise on Government, Wordsworth Classics.
**Louzek, Mark**. 2011. The Battle of Methods in Economics: The Classical Methodenstreit—Menger vs. Schmoller, The American Journal of Economics and Sociology, APRIL, 2011, Vol. 70, No. 2, Social, Methods, and Microeconomics: Contributions to Doing Economics Better (APRIL, 2011), pp. 439-463

**McConnell, Campbell R., Stanley L. Brue, and Sean M. Flynn**. 2018. Microeconomics: Principles, Problems, and Policies, McGraw-Hill.
**Milonni, Peter W**. 2004. Fast Light, Slow Light, and Left-Handed Light, CRC Press.
**Mises, Ludwig**. 1998. Human Action: A Treatise on Economics (Scholar's Edition), Ludwig von Mises Institute.
**Mises, Ludwig**. 2006. The Ultimate Foundation of Economic Science: An Essay on Method, Liberty Fund.
**Mises, Ludwig**. 2007. Bureaucracy, Liberty Fund.
**Mises, Ludwig**. 2008. Profit and Loss, Ludwig von Mises Institute.
**Mises, Ludwig**. 2009. Liberty and Property, Ludwig von Mises Institute.
**Mises, Ludwig**. 2010. Omnipotent Government: The Rise of the Total State and Total War, Ludwig von Mises Institute.
**Mises, Ludwig**. 2012. Economic Calculation in the Socialist Commonwealth, Ludwig von Mises Institute.
**Mises, Ludwig**. 2015. Socialism: An Economic and Sociological Analysis, Liberty Fund.
**Nozick, Robert**. 2013. Anarchy, State, and Utopia, Basic Books. [Original publication year, 1974.]
**Ostrom, Elinor**. 2009. Beyond Markets and States: Polycentric Governance of Complex Economic Systems, Nobel Prize Lecture.
**Palmer, Tom. G**. 1990. "Are Patents and Copyrights Morally Justified? The Philosophy of Property Rights and Ideal Objects," in "Symposium: Intellectual Property," Harvard Journal of Law & Public Policy 13, no. 3 (Summer 1990), p. 818.
**Penner, James**. 2000. The Idea of Property in Law, Oxford University Press.
**Plant, Arnold**. 1934A. "The Economic Theory Concerning Patents for Inventions," Economica, New Series, 1 (1), (February 1934): 30–51.
**Plant, Arnold**. 1934B. "The Economic Aspects of Copyright in Books," Economica, New Series, 1 (2), (May 1934): 167–195.
**Plantinga, Alvin C**. 1993A. Warrant: The Current Debate, Oxford University Press.
**Plantinga, Alvin C**. 1993B. Warrant and Proper Function, Oxford University Press.
**Radin, Margaret Jane**. 1982. "Property and Personhood." *Stanford Law Review* 34 (4): 957–1015.
**Rawls, John**. 1971. *A Theory of Justice*. Cambridge, MA: Belknap Press of Harvard University Press.
**Rawls, John**. 2001. Justice as Fairness: A Restatement, Erin Kelly (ed.), Belknap Press.
**Rose, Carol M**. 1996. Property as the Keystone Right?, 71 Notre Dame L. Rev. 239 (1996).
**Rothbard, Murray**. 1982. Law, Property Rights, and Air Pollution, Cato Journal, vol. 2, No. 1 (Spring 1982), pp. 55 - 99.
**Rothbard, Murray**. 1991. The End of Socialism and the Calculation Debate Revisited, The Review of Austrian Economics, Vol. 5, No. 2 (1991): 51-76.
**Rothbard, Murray**. 2002. The Ethics of Liberty, New York University Press.
**Rothbard, Murray**. 2005. For A New Liberty: The Libertarian Manifesto, 2nd ed., Ludwig von Mises Institute.
**Rothbard, Murray**. 2006. An Austrian Perspective on the History of Economic Thought, Vol. 1, Ludwig von Mises Institute.
**Rothbard, Murray**. 2009A. Anatomy of the State, Ludwig von Mises Institute.
**Rothbard, Murray**. 2009B. Man, Economy, and State, with Power and Market, Ludwig von Mises Institute.
**Rothbard, Murray**. 2011. Economic Controversies, Ludwig von Mises Institute.
**Rowe, M. W**. 1992. The Definition of 'Game,' Philosophy, Oct., 1992, Vol. 67, No. 262 (Oct., 1992), pp. 467-479.
**Samuelson, Paul A. and William D. Nordhaus**. 2010. Economics, 19th ed., McGraw-Hill.
**Sen, Amartya**. 1988. Property and Hunger, Economics and Philosophy, 4, 1988, 57-68.
**Shafer-Landau, Russ**. 2015. The Fundamentals of Ethics, 3rd ed., Oxford University Press.
**Shapiro, Stewart, and Craige Roberts**. 2019. Open Texture and Analyticity. From "Friedrich Waismann: The Open Texture of Analytic Philosophy," ed. by Dejan Makovec and Stewart Shapiro, Palgrave Macmillan.

**Sowell, Thomas**. 1996. Knowledge and Decisions, Basic Books.
**Spooner, Lysander**. 1884. A Letter to Scientists and Inventors, on the Science of Justice, and their Right of Perpetual Property in their Discoveries and Inventions, from The Shorter Works and Pamphlets of Lysander Spooner, Vol. 2 (1862-1884).
**Stake, Jeffrey Evans**. 2004. The Property 'Instinct,' Phil. Trans. R. Soc. Lond. B (2004) 359, 1763-1774, doi: 10. 1098/rstb.2004.1551.
**Waismann, Friedrich**. 1945. Symposium: "Verifiability". Proceedings of the Aristotelian Society (Supp.) 19, 119-50.
**Waldron, Jeremy**. 1979. Enough and as Good Left for Others, The Philosophical Quarterly (1950-), Vol. 29, No. 117 (Oct., 1979), pp. 319-328.
**Waldron, Jeremy**. 1983. Two Worries About Mixing One's Labour, The Philosophical Quarterly, Jan., 1983, Vol. 33, No. 130, pp. 37-44.
**Waldron, Jeremy**. 2023. Property and Ownership, the Stanford Encyclopedia of Philosophy.
**Weitz, Morris**. 1956. The Role of Theory in Aesthetics. The Journal of Aesthetics and Art Criticism, Sep., 1956, Vol. 15, No. 1, pp. 27 - 35.
**Wittgenstein, Ludwig**. 1991. Philosophical Investigations: The German Text, with a Revised English Translation 50th Anniversary Commemorative Edition, 3rd ed. Wiley-Blackwell.